\documentclass[aps,prb,amsmath,amssymb,nofootinbib,twocolumn,
superscriptaddress]{revtex4-1}

\usepackage{color,array}
\usepackage{slashed}
\usepackage{graphicx}
\usepackage{epsfig}
\usepackage{bm}
\usepackage{psfrag}
\usepackage{hyperref}
\newcommand{\dg}{\dagger}
\newcommand{\up}{\uparrow}
\newcommand{\down}{\downarrow}

\newcommand{\tp}{t_\perp}
\newcommand{\avg}[1]{\langle #1\rangle}
\newcommand{\vect}[1]{\boldsymbol{#1}}

\begin{document}

\title{Ground state phase diagram of the half-filled bilayer Hubbard model}
\date{\today}

\author{Michael Golor}
\email{golor@physik.rwth-aachen.de}
\affiliation{\mbox{\it Institut f\"ur Theoretische Festk\"orperphysik, JARA-FIT and JARA-HPC, RWTH Aachen University, 52056 Aachen, Germany}}

\author{Timo Reckling}
%\email{reckling@thphys.uni-heidelberg.de}
\affiliation{\mbox{\it 	Institut f\"ur Theoretische Physik, Universit\"at Heidelberg,}
\mbox{\it 69120 Heidelberg, Germany}}

\author{Laura Classen}
%\email{classen@thphys.uni-heidelberg.de}
\affiliation{\mbox{\it 	Institut f\"ur Theoretische Physik, Universit\"at Heidelberg,}
\mbox{\it 69120 Heidelberg, Germany}}

\author{Michael M. Scherer}
\email{scherer@thphys.uni-heidelberg.de}
\affiliation{\mbox{\it 	Institut f\"ur Theoretische Physik, Universit\"at Heidelberg,}
\mbox{\it 69120 Heidelberg, Germany}}

\author{Stefan Wessel}
%\email{wessel@physik.rwth-aachen.de}
\affiliation{\mbox{\it Institut f\"ur Theoretische Festk\"orperphysik, JARA-FIT and JARA-HPC, RWTH Aachen University, 52056 Aachen, Germany}}

%%%%%%%%%%%%%%%%%%%%%%%%%%%%%%%%%%
\begin{abstract}
Employing a combination of functional renormalization group calculations and projective determinantal quantum Monte Carlo simulations, we examine the Hubbard model on the  square lattice bilayer at half filling. 
From this combined analysis, we obtain a comprehensive account on the ground state phase diagram with respect to the extent of the system's metallic and (antiferromagnetically ordered)  Mott-insulating as well as band-insulating regions.
By means of an unbiased functional renormalization group approach, we exhibit the antiferromagnetic Mott-insulating state as the relevant instability of the free metallic state, induced by any weak finite onsite repulsion. 
Upon performing a careful analysis of the quantum Monte Carlo data, we resolve the difficulty of identifying this antiferromagnetic ground state for finite interlayer hopping in the weak-coupling regime, where nonmonotonous finite-size corrections are shown to relate to the two-sheeted Fermi surface structure of the metallic phase. 
On the other hand, quantum Monte Carlo simulations are well suited to identify the transition between the Mott-insulating phase and the band insulator in the intermediate-to-strong coupling regime. Here, we compare our numerical findings to indications for the transition region obtained from the functional renormalization group procedure.
\end{abstract}
%%%%%%%%%%%%%%%%%%%%%%%%%%%%%%%%%%

\maketitle

%%% Introduction %%%%%%%%%%%%%%%%%%%%%%%%
%%%%%%%%%%%%%%%%%%%%%%%%%%%%%%%%%%
\section{Introduction}
%%%%%%%%%%%%%%%%%%%%%%%%%%%%%%%%%%

The Hubbard model provides a fundamental description of correlation effects in many-body condensed matter systems. 
On the  square lattice bilayer, it exhibits several different routes for the transition out of normal metallic behavior by an independent tunability of (i) the free band-structure via a single tight-binding parameter, i.e., the interlayer hopping amplitude, (ii) the onsite repulsion, as well as (iii) the filling. 
For example, at half filling, the interlayer hopping controls a continuous transition from a Mott-insulating state to a band insulator~\cite{fuhrmann2006,kancharla2007,bouadim2008,hafermann2009,napitu2012,euverte2013,rademaker2013,rademaker2013b,capponi2004,ruger2014,lee2014}. 
Furthermore, at finite doping the model shows the appearance of superconducting pairing instabilities and by tuning the interlayer hopping, a transition between distinct types of pairing symmetries can be accomplished, cf. e.g. Ref.~\onlinecite{zhai2009}. 
These pairing symmetries, of $d$-wave and extended $s$-wave character, are frequently discussed in the context of the cuprate and pnictide compounds, respectively\cite{bulut1992,liechtenstein1995,kuroki2002,zhai2009,maier2011,cho2013}.
Finally, the bilayer Hubbard model was also discussed as a simple model with multiple Fermi surfaces and electron and hole pockets, cf. e.g. Ref.~\onlinecite{shimizu2007}, as well as in the context of copper oxide bilayers featuring high-$T_c$ superconductivity~\cite{damascelli2003,fournier2010}. 
The bilayer Hubbard model might as well become a rich and instructive example system with a well-defined and at the same time complex phase diagram, given the progress in the controlled manipulation of ultracold atoms in optical lattices~\cite{esslinger2010}.

The bilayer Hubbard model at half filling has been studied independently using various methods recently, ranging from dynamical mean field theory (DMFT) and cluster extensions\cite{fuhrmann2006,kancharla2007,hafermann2009,napitu2012} and variational Monte Carlo\cite{ruger2014}, to finite-temperature determinantal quantum Monte Carlo (DQMC) studies\cite{scalettar1994,bouadim2008,rademaker2013}. 
At present, however, it appears that some of the results on its phase diagram remain inconclusive -- especially since the  persistence of an extended paramagnetic metallic phase at small onsite interactions has been put forward based on cluster DMFT\cite{kancharla2007} as well as finite temperature DQMC calculations\cite{bouadim2008} -- a result, which contrasts  the perfect nesting property of the noninteracting system at half filling throughout the whole metallic regime. 
The above outlined perspectives for the physics of the bilayer Hubbard model hence call for a clarification concerning its ground state phase diagram at half filling.

Here, we  provide a comprehensive account on  the ground state phase diagram by combining complementary methods, which -- taken together -- allow us to cover the full range of the local interaction strength.
For this purpose, we first review the standard Hartree-Fock mean-field theory (HFMFT) approximation to properly settle the mean-field character of the antiferromagnetic state as a function of the 
onsite interaction, as arising from a Stoner instability of the metallic noninteracting state. 
In a second step, which leads systematically beyond mean-field theory as well as random phase approximations (RPA), we implement a functional renormalization group (fRG) method in the discrete patching scheme, which provides a well-controlled approach at weak interactions and allows us to obtain reliable results for intermediate interaction strengths~\cite{metzner2011,Honerkamp2008,platt2013}. 
An important benefit provided by this method is that it takes into account effects from possibly competing correlations and therefore allows us to detect the appearance of instabilities in an 
unbiased way, i.e. without a priori assumptions concerning the nature of the emerging order. 
Finally, we employ an unbiased and numerically exact method, zero-temperature (projective) DQMC, which is particularly powerful in the regime of intermediate to strong coupling and which allows us to identify actual ground state correlations on finite systems. 
It turns out that especially the mutual strengths of these  approaches help to provide a coherent picture of the low-energy physics of the bilayer Hubbard model.

The rest of this paper is organized as follows: In Sec.~\ref{sec:model-phase}, we introduce the Hubbard model on the square lattice bilayer and also summarize our central findings from the subsequent sections by presenting the resulting ground state phase diagram of the half-filled system. 
Sec.~\ref{sec:weak} provides a detailed account on the weak-coupling regime from the perspective of three different many-body methods: HFMFT, fRG, and DQMC. 
Our projective DQMC calculations exhibit rather severe, nonmonotonous finite-size effects in the weak-coupling regime, that most likely explain  conflicting previous conclusions on the persistence 
of the metallic state drawn from finite-temperature DQMC simulations~\cite{bouadim2008} (which also require control of the simulation temperature appropriately in order to sample  ground state correlations). 
In Sec.~\ref{sec:afbi}, we examine the transition from the antiferromagnetic Mott insulator to the strong-interlayer-hopping band insulator and contrast its identification within the DQMC and the fRG approach. 
Final conclusions will be presented in Sec.~\ref{concl}.

%%% Model %%%%%%%%%%%%%%%%%%%%%%%%%%%
\section{Model and Phase Diagram}
\label{sec:model-phase}

In  the following, we examine the Hubbard model on the square lattice bilayer with intra-layer nearest-neighbor hopping $t$, inter-layer hopping $t_\perp$ and a local Coulomb repulsion $U$, as described by the Hamiltonian
\begin{align}
\label{eq:model}
	H= &-t \sum_{\langle ij \rangle s \lambda}\!\!\left(c_{i \lambda s}^{\dagger}c^{}_{j \lambda 		s} + \mathrm{h.c.}\right) \\
 &-t_{\perp} \sum_{is}{\left(c_{i 1 s}^{\dagger}c^{}_{i 2 s} + \mathrm{h. c.}\right)} + U\sum_{i\lambda}n_{i\lambda\uparrow}n_{i\lambda\downarrow},  \nonumber
\end{align}
where the $c_{i \lambda s}^{(\dagger)}$ denote annihilation (creation) operators for electrons of spin $s\in\{\uparrow,\downarrow\}$ on site $i$ in layer $\lambda\in\{1,2\}$, and the $n_{i \lambda s}$ denote local occupation number operators.

\begin{figure}[b]
  \centering
  \includegraphics[width=\columnwidth]{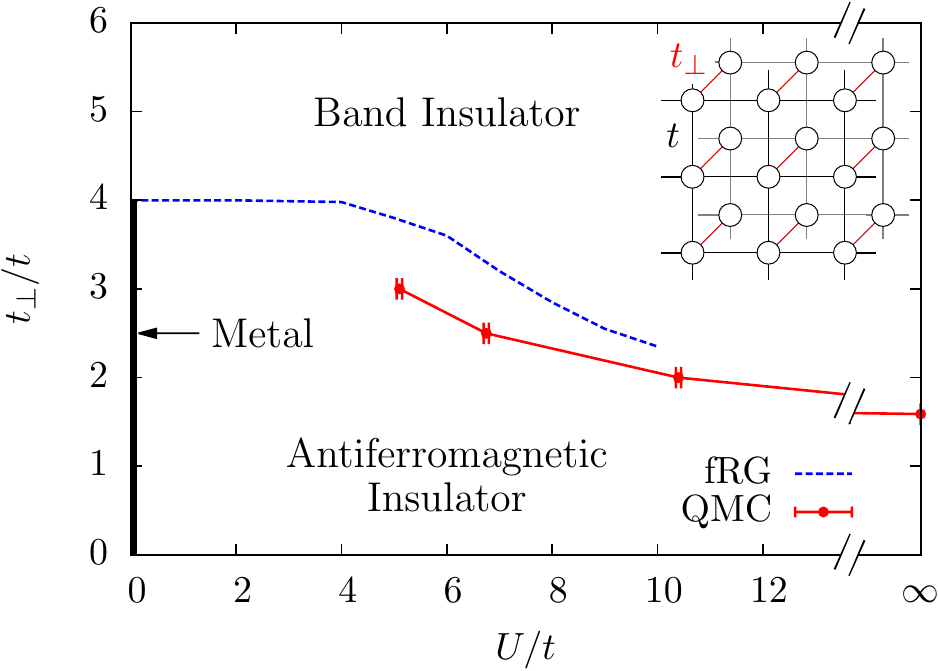}
  \caption{(color online) Phase diagram of the Hubbard model on the square lattice bilayer (inset) at half filling. Red dots display the phase boundary obtained from projective DQMC (the red solid line is a guide to the eye). The blue dashed line indicates the reduction of the sharply pronounced antiferromagnetic momentum structure upon increasing $t_\perp$ at intermediate values of $U/t$ in the fRG approach, hinting towards a phase boundary of the antiferromagnetic instability (cf. Sec.~\ref{sec:frg-afbi} for details).}
  \label{fig:phase-diag}
\end{figure}

At half-filling, our calculations reveal the phase diagram shown in Fig.~\ref{fig:phase-diag}.
The metallic phase only persists in the noninteracting case ($U=0$) for interlayer hoppings $t_\perp/t<4$. 
For larger values of $t_\perp/t$, the system becomes insulating due to the opening of a band gap.
In the weak coupling regime ($U\lesssim 4t$), the system is an antiferromagnetic insulator state, triggered by a Stoner instability, which is due to the perfect nesting property of the two-sheeted Fermi surface at wave vector $\bm Q=(\pi/a,\pi/a)$ with the lattice constant set to $a=1$ in the following, see Fig.~\ref{fig:fermi-surf}. The phase transition to the band insulator still occurs at $t_\perp\approx4t$.
For larger couplings $U/t$, this phase boundary is continuously shifted to $t_\perp=1.588t$ in the $U\to\infty$ limit: 
in this regime, the low-energy spin physics can be mapped onto that of a spin-$1/2$ Heisenberg model, which undergoes a quantum  phase transition
from an antiferromagnetic state to a dimerized phase with spin singlets predominantly formed on the interlayer bonds\cite{hida1992,millis1993,sandvik1994,wang2006}.
The following sections provide details of our analytical and numerical findings on which the identification of the phase diagram is based.

%%%%%%%%%%%%%%%%%%%%%%%%%%%%%%%%%%
\section{Weak-coupling regime}
\label{sec:weak}
%%%%%%%%%%%%%%%%%%%%%%%%%%%%%%%%%%

%%% Tight-Binding und MFT
%%%%%%%%%%%%%%%%%%%%%%%%%%%%%%%%%%
\subsection{Tight-Binding and Hartree-Fock Approximation}
\label{sec:tb-hf}
%%%%%%%%%%%%%%%%%%%%%%%%%%%%%%%%%%

\begin{figure*}[t]
  \centering
\includegraphics[width=0.9\textwidth]{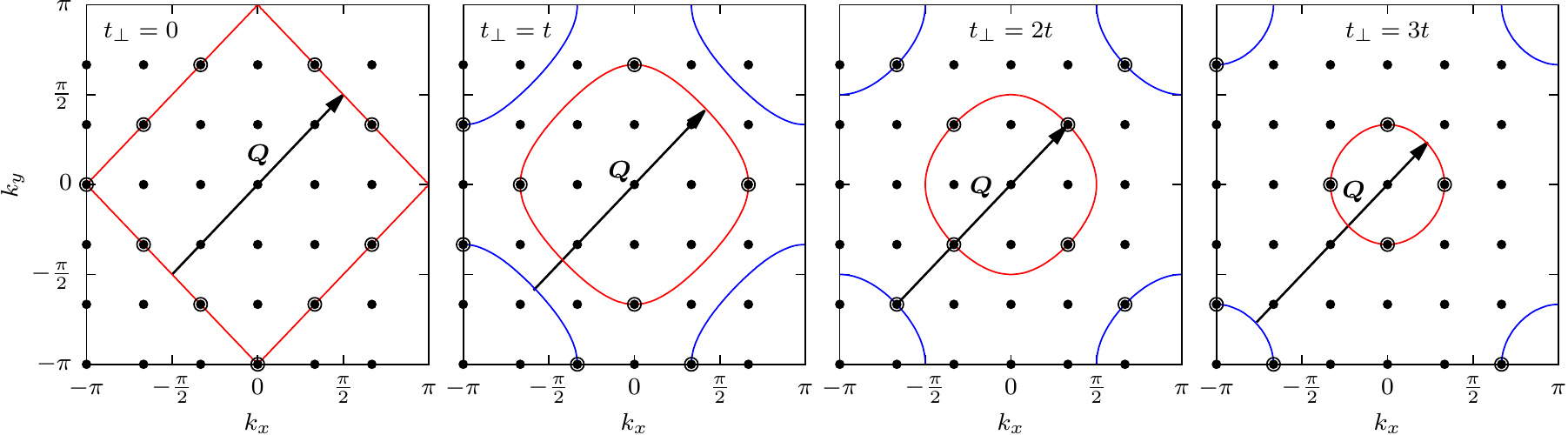}  
  \caption{(color online) Fermi surfaces of the square lattice bilayer tight-binding model ($U=0$) at half filling for different values of $t_\perp/t=0,1,2,3$ (from left to right). Every point in one band is connected to the other band by the nesting vector $\bm Q=(\pi,\pi)$. The  grid points denote the momenta included in a finite system of linear length $L=6$ with periodic boundary conditions. For the system sizes accessible in the DQMC simulations, these represent the only values of $t_\perp/t$ with discrete momenta located on the Fermi surface (larger dots).}
  \label{fig:fermi-surf}
\end{figure*}

Since the physics in the weak-coupling regime is strongly influenced by the structure of the Fermi surface of the noninteracting
system, we start our analysis by briefly reviewing this case. 
For $U=0$, Eq.~(\ref{eq:model}) reduces to a tight-binding model, which can be solved by exact diagonalization and 
exhibits the single-particle dispersion
\begin{equation}
\label{eq:tb-disp}
\epsilon^\pm_0(\vect{k},t_\perp) = -2t\left(\cos(k_x ) + \cos(k_y )\right)\pm t_\perp.
\end{equation}
This corresponds to two copies of the square lattice single-layer dispersion ($t_\perp=0$), shifted with respect to each other linearly in $t_\perp$.  
Since for the single-layer lattice the energies lie within the range $-4t \leq \epsilon\leq4t$, a band gap opens for $t_\perp>4t$ and the system becomes band-insulating at half filling.
The Fermi surface for various values of $t_\perp/t$ is shown in Fig.~\ref{fig:fermi-surf}. Both bands remain perfectly nested for all values $t_\perp/t$, with a nesting vector
$\vect{Q} = (\pi, \pi)$, i.e.
\begin{equation}
\epsilon^{+}_0(\vect{k} +\vect{Q},t_\perp) = -\epsilon^{-}_0(\vect{k},t_\perp).
\end{equation}

\begin{figure}[t]
\centering
%  \includegraphics[height=0.165\textheight]{tperp2_1BZ.pdf}
%  \hspace{0.2cm}
  \includegraphics[width=0.8\columnwidth]{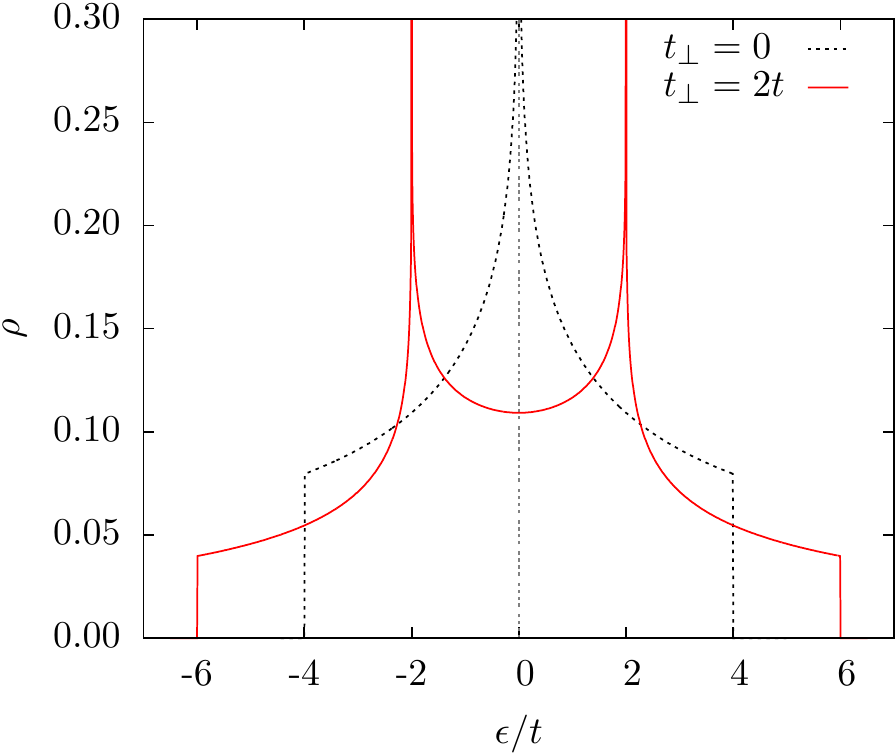}
  \caption{(color online) The $U=0$ bilayer density of states (DOS) $\rho(\epsilon, t_\perp)$ for $t_\perp = 2 t$ (solid red line) in comparison with the single-layer DOS (dashed black line). The dotted vertical line marks the position of the Fermi level.}
  \label{fig:tperp2_1BZ}
\end{figure}

As a consequence of Eq.~(\ref{eq:tb-disp}), the density of states (DOS) of the bilayer lattice, $\rho(\epsilon, t_\perp)$, is the sum of two displaced single-layer DOS contributions, see Fig.~\ref{fig:tperp2_1BZ}. 
This causes the logarithmic van Hove singularity of the single-layer square lattice  to be shifted away from the Fermi level to energies $\pm t_\perp$. 
As a consequence, one expects a suppression of interaction effects in the bilayer system.  However, for $\tp<4t$, the DOS stays finite at the Fermi surface. This, combined with the nesting property, leads to a divergence of the zero-temperature static spin-spin susceptibility 
\begin{equation}
  \label{eq:suscep}
  \chi_0^{+-}(\bm q)=-\frac1N\sum_{\bm k}\frac{f\left(\epsilon^+_0(\bm k+\bm q,\tp)\right)-f\left(\epsilon^-_0(\bm k,\tp)\right)}{\epsilon^+_0(\bm k+\bm q,\tp)-\epsilon^-_0(\bm k,\tp)},
\end{equation}                  
where $f(\epsilon)$ denotes the Fermi-Dirac distribution. At the nesting vector, $\chi_0^{+-}(\bm Q)\to\infty$ (for temperature $T\to 0$), rendering the system unstable towards antiferromagnetic order. 

A first access to the weakly interacting system may be obtained upon treating the interaction term in a self-consistent HFMFT approximation, i.e.,~after decoupling the interaction term 
\begin{equation}
  \label{eq:hf-decoupling}
  n_{i\lambda\up}n_{i\lambda\down}\rightarrow n_{i\lambda\up}\avg{n_{i\lambda\down}}+\avg{n_{i\lambda\up}}n_{i\lambda\down} -\avg{ n_{i\lambda\up}}\avg{ n_{i\lambda\down}}
\end{equation}
and expressing the occupation expectation values in terms of a homogeneous staggered magnetization $m=|\avg{n_{i\lambda\up}}-\avg{n_{i\lambda\down}}|/2$. The resulting effective Hamiltonian
can again be solved analytically (see App.~\ref{sec:hubb-mf} for details) and the dispersion is now given by four bands labeled by an index $\alpha\in\{1,...,4\}$,
\begin{align}
\label{eq:hf-disp}
\epsilon^\alpha_{\rm HF}(\vect{k},t_\perp) = \pm\sqrt{\left(\epsilon^\pm_0(\bm k, t_\perp)\right)^2+\Delta^2},\quad \Delta\propto t\,e^{-C_{\tp} t/U},
\end{align}
with a positive coefficient $C_{\tp}$ that increases with $\tp/t$.
Here, $\Delta=Um$ corresponds to the energy gap at the Fermi surface and decays exponentially with decreasing coupling strength $U/t$.
Nonetheless, the gap and the staggered magnetization stay finite for any finite interaction, indicating the onset of antiferromagnetism
for the weakly interacting system. Previous finite-temperature DQMC simulations\cite{bouadim2008} concluded in favor of a finite extent of the paramagnetic, metallic region for finite $t_\perp$, arguing that the tendency of the interlayer coupling towards the formation of spin-singlet states, which indeed prevails in the large-$U$ region, might compete with the HFMFT antiferromagnetic instability also at low values of $U/t$. As will be shown below, this conclusion  reflects an inherent difficulty of performing a systematic finite-size analysis of DQMC data in the weak-coupling region, due to nonmonotonous finite-size effects, even when ground state expectation values are accessed.
In the following section, we first employ an fRG approach that shows in an unbiased way how the antiferromagnetic instability indeed emerges in the weak-interaction regime from the noninteracting metallic state.  

%%% FRG-Teil
%%%%%%%%%%%%%%%%%%%%%%%%%%%%%%%%%%
\subsection{Functional Renormalization Group}
%%%%%%%%%%%%%%%%%%%%%%%%%%%%%%%%%%
\label{sec:frg_weak}

\begin{figure}[t!]
\centering
	\includegraphics[height=0.5\columnwidth]{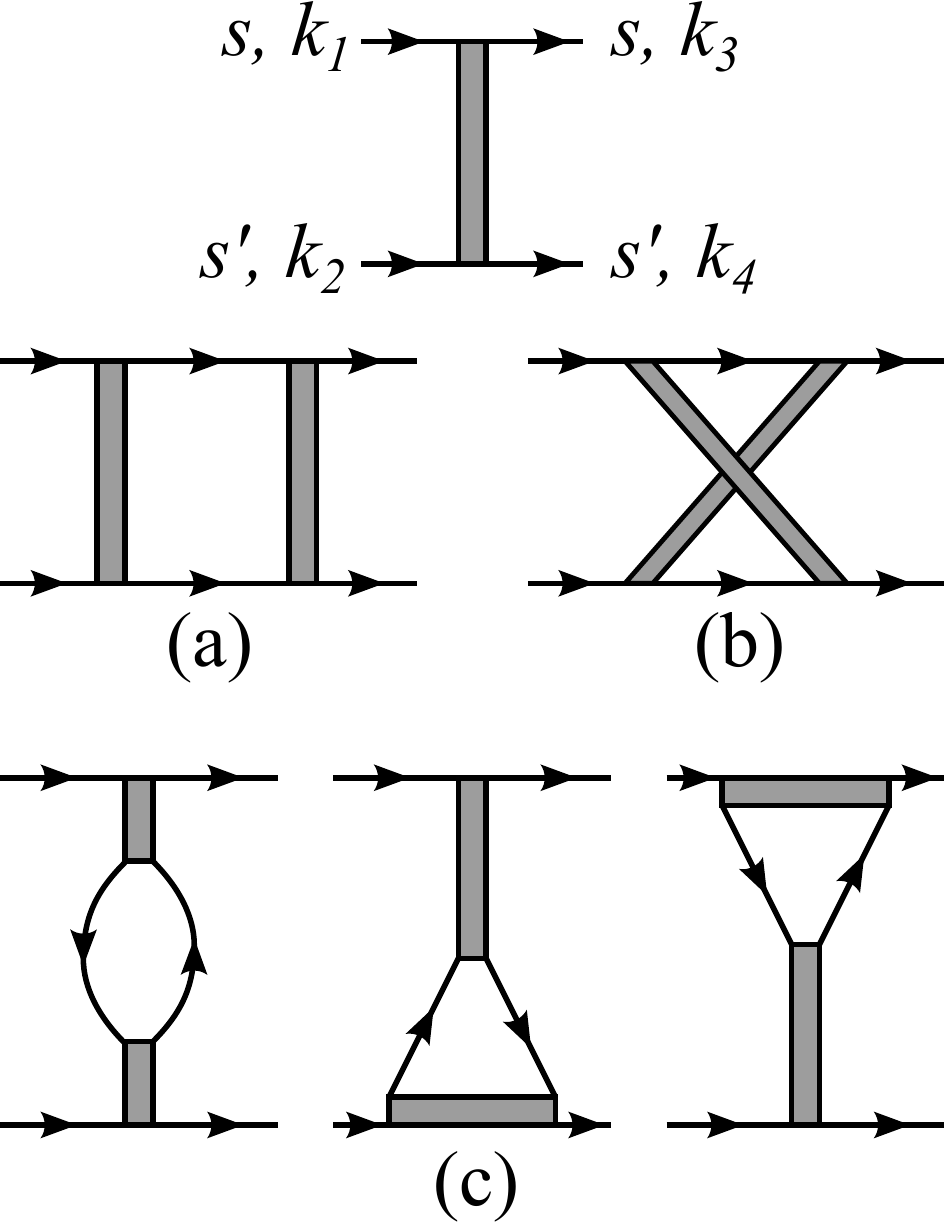}
	\hspace{0.2cm}
  	\includegraphics[height=0.5\columnwidth]{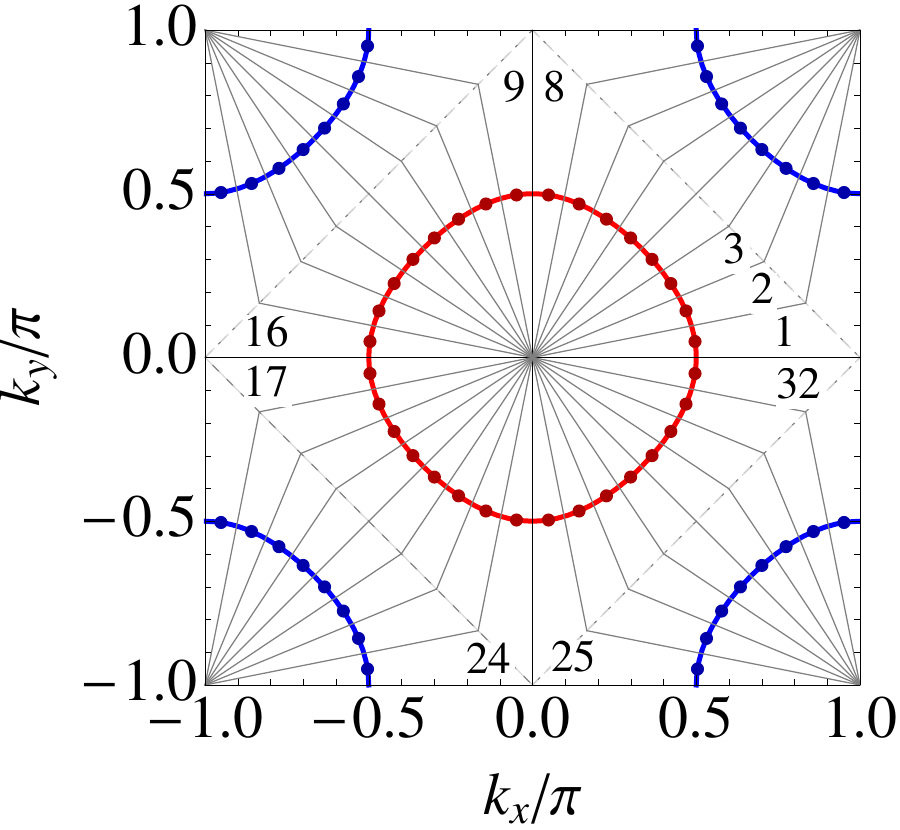}
  	\caption{(color online) Left panel on top: Interaction vertex and spin convention. Below: Loop contributions from the particle-particle channel (a), the crossed particle-hole channel (b) and the direct particle-hole channel (c). Right panel: Patching scheme of the Brillouin zone for a total of $N_p=32$ patches. Depending on the energy band the coupling function is evaluated on a wave vector at the Fermi level indicated by the dots (red: upper band, blue: lower band).}
  \label{fig:patching}
\end{figure}

Here, we employ the fRG method for the one-particle-irreducible (1PI) vertices of a fermionic many-body system to obtain their renormalization group (RG) evolution for the bilayer  system, see e.g.~Refs.~\onlinecite{metzner2011,platt2013} for details on the general fRG approach. 
To that end, the bare propagator of the action corresponding to the model defined in Eq.~\eqref{eq:model} is modified by introducing an infrared regulator with energy scale $\Lambda$.
Varying $\Lambda$ generates the RG flow in terms of a functional differential equation. 
The flow is then integrated out starting with the bare action at the initial scale $\Lambda_0$ corresponding to the bandwidth of the system.
We integrate towards the infrared limit $\Lambda \rightarrow 0$,  which induces a smooth interpolation between the bare action of the system and the effective action at low energies.
The fRG equations obtained from this procedure amount to an infinite hierarchy of coupled flow equations for the 1PI vertex functions and approximations are in order for their practical evaluation.
In our numerical implementation we employ a truncation, where we follow the flow of the two-particle interaction vertex described by the coupling function $V_\Lambda(k_1,k_2;k_3,k_4)$ where $k_i$ consists of a wavevector $\vect{k}_i$, a Matsubara frequency $\omega_i$, a spin projection $s_i$, and band indices $b_i$ or layer indices $\lambda_i$.
External frequencies are set to zero to study the ground state and furthermore,  the momentum dependence is discretized.
We also neglect selfenergy corrections to reduce the numerical effort. 
The vertex $V_\Lambda$ is discretized by dividing the Brillouin zone into $N_p$ patches with a constant wavevector dependence within each patch, as shown in Fig.~\ref{fig:patching}.
Representative momenta for these patches are chosen to reside at the Fermi level.
In the following, we employ a wavevector resolution with a patch number of $N_p=32$ and $N_p=48$. We have also checked our results to be stable towards higher resolutions with up to $N_p=96$ patches. 
In summary, we obtain a vertex function $V_\Lambda$ with $N_p^3\cdot N_b^4$ components, where $N_b=2$ is the number of energy bands, and a set of $N_p^3\cdot N_b^4$ coupled differential equations that has to be integrated.

The approximations involved in this approach correspond to an infinite-order summation of one-loop particle-particle and particle-hole diagrams, cf. Fig.~\ref{fig:patching}, which allows the study of various competing correlation effects. 
Eventually, this leads to instabilities in the RG flow at a critical scale $\Lambda_c$, where components of the vertex grow large.
Such a divergence of the interaction vertex can be seen as an artifact of our truncation due to the approximations described above.
On the other hand a divergence in the interaction vertex is a dependable indicator for a tendency towards the formation of an ordered state and the well-pronounced emerging momentum structure close to the critical scale allows us to identify an effective Hamiltonian for the low-energy regime.
In practice, the flow is halted at a finite value for the largest vertex component of the order of several times the bandwidth. 
This defines an energy scale which gives a reasonable approximation to the critical scale as the vertex diverges quickly close to the instability\cite{metzner2011,honerkamp2001}.
Near the scale  $\Lambda_c$, the dominant correlations become clearly detectable in the vertex function and we can use this scale as an estimate for e.g. the energy scale below which the single-particle spectrum gets modified, e.g. by a gap.
The procedure described here is well-controlled for small interactions and can be expected to be reliable also in the regime of intermediate interaction strengths~\cite{metzner2011,platt2013,eberlein2013}.
Importantly, the fRG in this approximation scheme takes into account effects from competing correlations and allows to identify the leading instabilities in an unbiased way, i.e., no a priori assumptions concerning the nature of the appearing order are required. The effect of the inclusion of self-energy feedback and frequency dependent interactions has been studied in the single-layer Hubbard model in Ref.~\onlinecite{uebelacker2012} where it has been shown that the leading channel is not strongly affected by this extension of the truncation.

\begin{figure}[t!]
\centering
   \includegraphics[width=1.0\columnwidth]{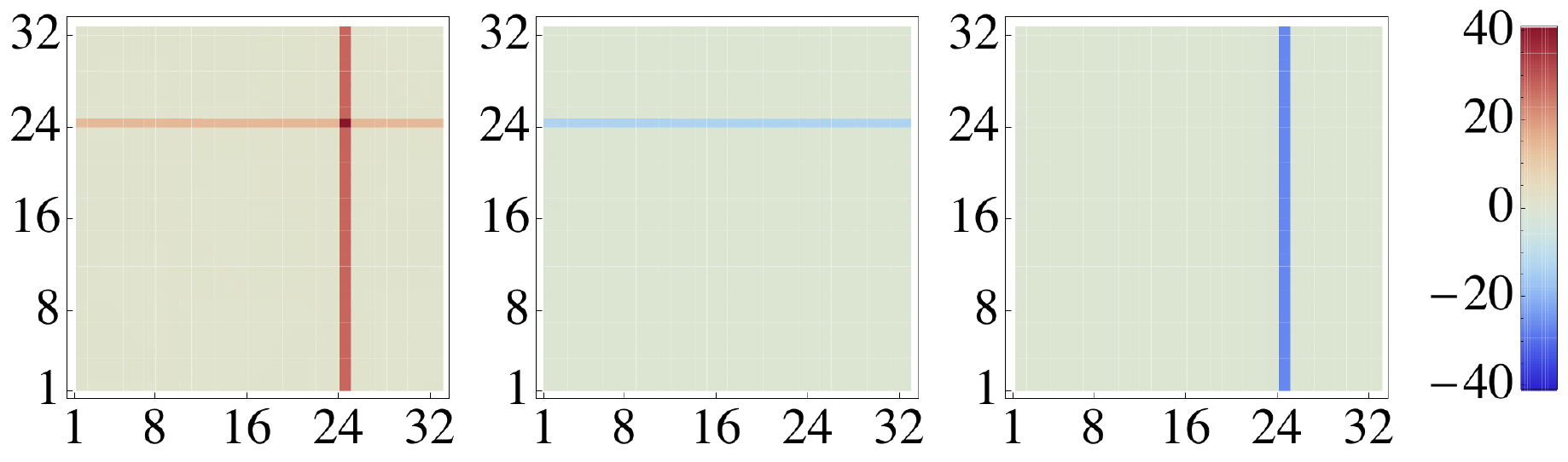}
  \caption{(color online) Effective interaction vertex near the critical scale for the AF-SDW in units of $t$, exhibiting a sharply pronounced momentum structure. The axes are numbered according to the number of the patch, cf. Fig.~\ref{fig:patching}, where wavevectors $k_1$ are depicted vertically and $k_2$ are depicted horizontally. We fix $k_3$ to be on patch 1. Left Panel: Effective vertex or a combination of layer indices $\lambda_i$ where $\lambda_1=\lambda_2=\lambda_3=\lambda_4$. Middle Panel: $\lambda_1=\lambda_3, \lambda_2=\lambda_4$. Right panel: $\lambda_1=\lambda_4, \lambda_2=\lambda_3$.}
  \label{fig:fRGvertex}
\end{figure}

In the fRG data at half filling and for arbitrary onsite repulsion, we observe as the prevailing  divergence an antiferromagnetic spin density wave (AF-SDW) with momentum transfer $\vect{Q}=(\pi,\pi)$, see Fig.~\ref{fig:fRGvertex} for a snapshot of the four-Fermi vertex close to the critical scale. 
This behavior clearly reflects the perfect nesting of the Fermi surface. The leading part of the interaction close to the critical scale can be expressed in terms of an effective interaction Hamiltonian
\begin{equation}
H_{\mathrm{SDW}}=-J\sum_{\lambda,i,j}e^{i {\bf Q}\cdot (\vect{r}_i-\vect{r}_j)}\big(\vect{S}_{i\lambda}\cdot\vect{S}_{j\lambda}-\vect{S}_{i\lambda}\cdot\vect{S}_{j\bar\lambda}\big)\,,
\end{equation}
where $J>0$, and $\bar \lambda$ denotes the layer opposite to $\lambda$.
The spin operator $\vect{S}_{i\lambda}$ is given by the relation $\vect{S}_{i\lambda}=1/2\sum_{s,s'}c_{i\lambda s}^\dagger \vect{\sigma}_{ss'}c_{i\lambda s'}$ in terms of the Pauli matrices. As a result of the sharpness in momentum space the interaction becomes long-ranged in position space. With this effective Hamiltonian, we can thus perform a controlled mean-field decoupling. 
For the resulting effective Hamiltonian, the spins are aligned antiferromagnetically within each layer and also between the layers. 

\begin{figure}[t!]
\centering
  \includegraphics[width=1.0\columnwidth]{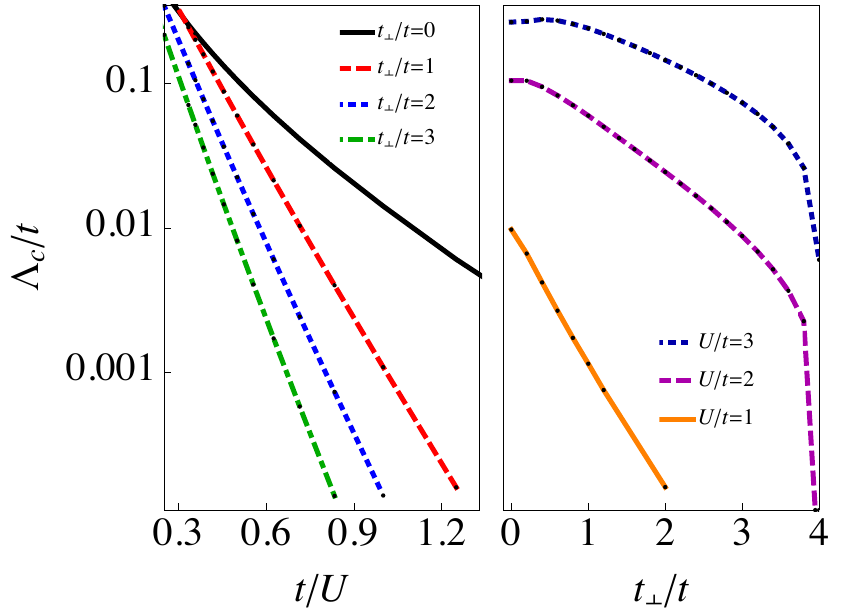}
  \caption{(color online) Left panel: fRG critical scale $\Lambda_c$ as a function of $t/U$ for the single-layer case (solid line) and the bilayer case with $t_{\perp}=1t,2t,3t$ (from top to bottom) using $N_p=48$ patches, exhibiting exponential decrease of $\Lambda_c\sim e^{-\mathrm{const.}\cdot t/U}$. This functional dependence suggests an instability for any value of $U>0$ in accord with HFMFT. Right panel: The fRG critical scale $\Lambda_c$ decreases by several orders of magnitude when $t_{\perp}/t$ is increased from $0$ to $4$. We show this behavior for three choices of $U/t=3,U/t=2$ and $U/t=1$ by the solid, dashed and dotted lines, respectively.}
  \label{frgUvar}
\end{figure}

In Fig.~\ref{frgUvar}, we compare the critical scales for several cases, namely the situation with vanishing perpendicular hopping amplitude $t_\perp=0$ which is equivalent to the one-band Hubbard model as well as for perpendicular hoppings of $t_\perp/t=1,2,3$ to compare with other methods, see also the caption of Fig.~\ref{frgUvar}. 
For the single-layer case, we observe that in agreement with the expectations from HFMFT the functional dependence of the critical scale on the onsite interaction $U/t$ is an exponential $\sim e^{-\mathrm{const.}\cdot\sqrt{t/U}}$, cf. Fig.~\ref{frgUvar}. 
In contrast, the critical scales for the bilayer case are considerably lower, which can be understood given the reduced density of states at the Fermi level.
Furthermore, the functional dependence on $U/t$ is different and rather follows a behavior $\sim e^{-\mathrm{const.}\cdot t/U}$ which is in accordance with the HFMFT expectation (cf. Sec.~\ref{sec:tb-hf}). 
Using the fRG approach, treating all the fluctuation channels on equal footing, we can also check whether  competing instabilities are present for this choice of parameters. 
Here, we do not observe the appearance of any other strong correlation effects apart from the AF-SDW instability. 
This also constitutes a worthwhile cross-check for the DQMC approach, excluding competing correlations as possible contributions to the observed finite-size scaling (cf. Sec.~\ref{qmc}).

In the right panel of Fig.~\ref{frgUvar} we compare the $t_\perp$-dependence of the critical scale $\Lambda_c$ for different values of $U/t \in \{1,2,3\}$. 
We observe a continuous decrease of $\Lambda_c$ as a function of increasing interlayer hopping, in accord with the HFMFT approximation, cf. App.~\ref{sec:hubb-mf}.

%%% QMC-Teil
%%%%%%%%%%%%%%%%%%%%%%%%%%%%%%%%%%
\subsection{Quantum Monte Carlo}
\label{qmc}
%%%%%%%%%%%%%%%%%%%%%%%%%%%%%%%%%%
We next employ projective ($T=0$) DQMC simulations to study the ground state correlations on finite systems. 
The DQMC approach enables us to obtain ground state expectation values of an arbitrary observable $O$ by projecting a trial wave function $|\Psi_\text{T}\rangle$ to the interacting ground state,
\begin{equation}
 \frac{\langle \Psi_0 | O | \Psi_0\rangle}{\langle \Psi_0 | \Psi_0\rangle}=\lim_{\Theta\to\infty} \frac{\langle \Psi_\text{T} | e^{-\Theta H} \,O\, e^{-\Theta H}| \Psi_\text{T}\rangle}{\langle \Psi_\text{T} |e^{-2\Theta H}| \Psi_\text{T}\rangle} \;.
\end{equation}
Here, we take the ground state of the free system ($U=0$) as the trial wave function in all simulations. Possible degeneracies are lifted by a slight dimerization\cite{hohenadler2012} of the intralayer hopping amplitude $t$ of the tight-binding system ($\delta t /t =10^{-4}$).
The projection parameter $\Theta$ has to be chosen sufficiently large, such that convergence to the ground state $|\Psi_0\rangle$ is guaranteed.
Depending mostly on the linear system size $L$ and the coupling strength $U/t$, values of $\Theta$ between $\Theta=30/t$ and $120/t$ are required to ensure convergence.
We employ a symmetric Suzuki-Trotter decomposition with an imaginary-time discretization of $\Delta\tau=0.1/t$, such that discretization errors are well below the size of the statistical errors. 
The Hubbard interaction is decoupled by a $SU(2)$ symmetric Hubbard-Stratonovich transformation, so that spin rotational symmetry is preserved during the entire simulation. For a detailed account of the projective DQMC algorithm, see Ref.~\onlinecite{assaad2008}.
The available computational resources allow us to simulate systems with linear lengths of up to $L=24$, i.e.~lattices containing up to $N=2\cdot L^2=1152$ sites.
In all cases, periodic boundary conditions in both dimensions are applied, ensuring translational symmetry.

In the weak-coupling regime the physics will be dominated by the structure of the Fermi surface and its nesting properties in particular. The finite systems with periodic boundary conditions, treated within DQMC, translate in momentum space to a finite set of $\bm k$-points sampling the Brillouin zone.
In order to include the relevant low-energy processes, we need to ensure that this set contains points at the Fermi surface.
Due to the specific Fermi surface structure of the bilayer square lattice, this considerably restricts our choice of $\tp$ and $L$, see Fig.~\ref{fig:fermi-surf}.
In particular, only for interlayer hopping $t_\perp/t\in\{1,2,3\}$, there are several system lengths $L\leq24$ available such as to allow for at least a limited finite-size analysis. 

\begin{figure}[t!]
  \centering
\includegraphics[width=\columnwidth]{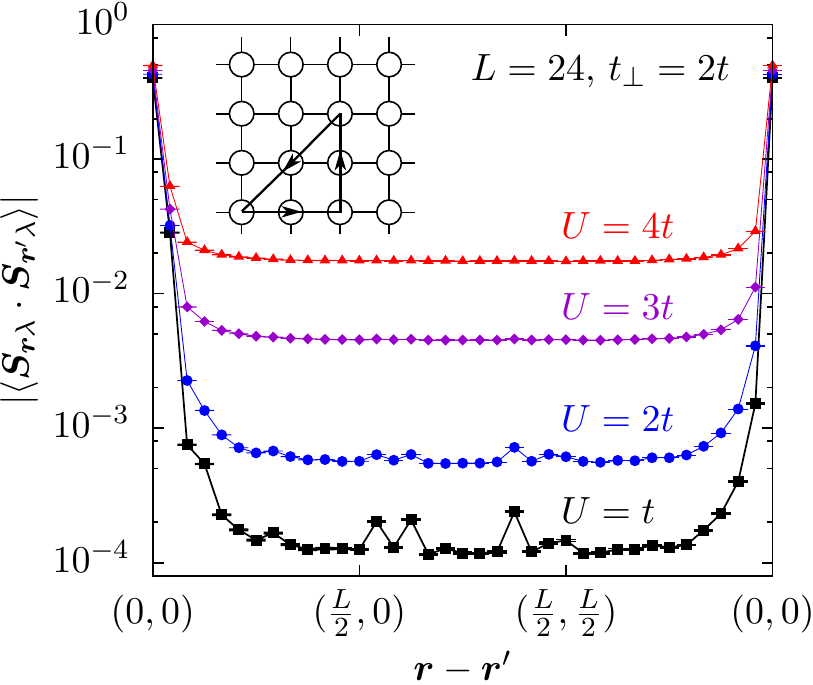}
  \caption{(color online) Static spin-spin correlations along a triangular path (inset) for different coupling strengths $U/t$. The calculations were performed for a bilayer system with interlayer hopping $\tp=2t$  and linear length $L=24$. }  
\label{fig:spinspin}
\end{figure}

The DQMC method allows for a direct access to static spin-spin correlations $\langle\bm S_{\bm r \lambda}\cdot\bm S_{\bm r' \lambda'}\rangle$ between two sites at positions $\bm r,\bm r'$ in layers $\lambda,\lambda'$. For the bilayer lattice with $\tp=2t$ and $L=24$, the spin-spin correlations along a triangular path within one layer of the lattice are shown in Fig.~\ref{fig:spinspin}. With increasing distance, the correlations decay quickly into the
long-distance regime. The magnitude of the asymptotic value falls off drastically upon reducing $U$ and for the lower values of $U$, the curves exhibit characteristic anomalies, which already hint towards the difficulty of extrapolating the finite-size data to the thermodynamic limit that will be considered in more detail, below.

A spin density wave (SDW) state with ordering vector $\bm q$ can be identified by monitoring the structure factor $S^{\pm}(\bm q)$, which is obtained by a Fourier transform of the measured spin-spin correlations, 
\begin{equation}
  \label{eq:sfd}
  S^\pm(\bm q)=\frac1N\sum_{\bm r,\bm r',\lambda} e^{i\bm q\cdot(\bm r-\bm r')}\left[\langle\bm S_{\bm r \lambda}\cdot\bm S_{\bm r' \lambda}\rangle\pm\langle\bm S_{\bm r \lambda}\cdot\bm S_{\bm r' \bar\lambda}\rangle\right],
\end{equation}
where $\bar \lambda$ denotes the layer  opposite to $\lambda$.
Due of the perfect nesting and the bipartite nature of the bilayer lattice, we expect an antiferromagnetic (AF) ordering, i.e.~a SDW with ordering vector $\bm Q=(\pi,\pi)$ and the corresponding antisymmetric structure factor $S^{\rm AF}=S^-(\bm Q)$.

The order parameter for the AF state, the staggered magnetization $m$, is then usually estimated by $\bar m=\sqrt{S^{\rm AF}/N}$. 
However, this estimator is disadvantageous in situations where the long distance correlations are of very low magnitude, as  is the case in Fig.~\ref{fig:spinspin} for small values of $U/t$.
The AF structure factor will then be dominated by local correlations, which are not suited to characterize a long-range ordered antiferromagnetic state on finite lattices. 
We therefore restrict the summation in Eq.~(\ref{eq:sfd}) to sites $\bm r, \bm r'$ of distance $|\bm r-\bm r'|>L/4$, where in all considered cases the correlations have decayed to their asymptotic level. The prefactor $1/N$ is modified accordingly. We denote this structure factor by $S^{\rm AF}_{L/4}$ and the corresponding estimator for the staggered magnetization by $\bar m_{L/4}$.

\begin{figure}[t!]
  \centering
\includegraphics[width=\columnwidth]{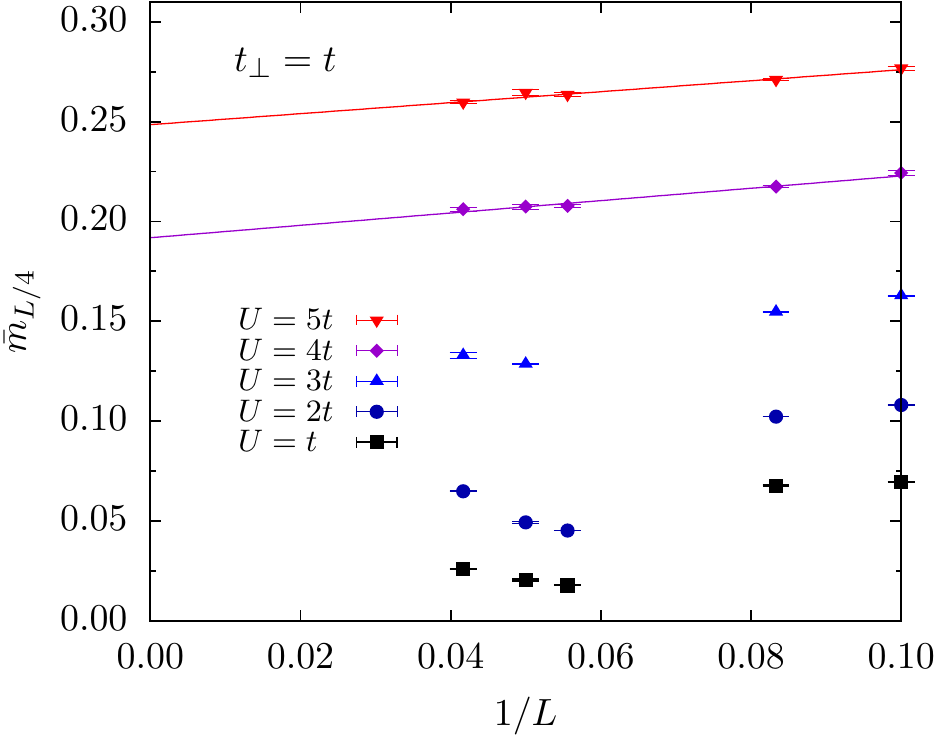}
\includegraphics[width=\columnwidth]{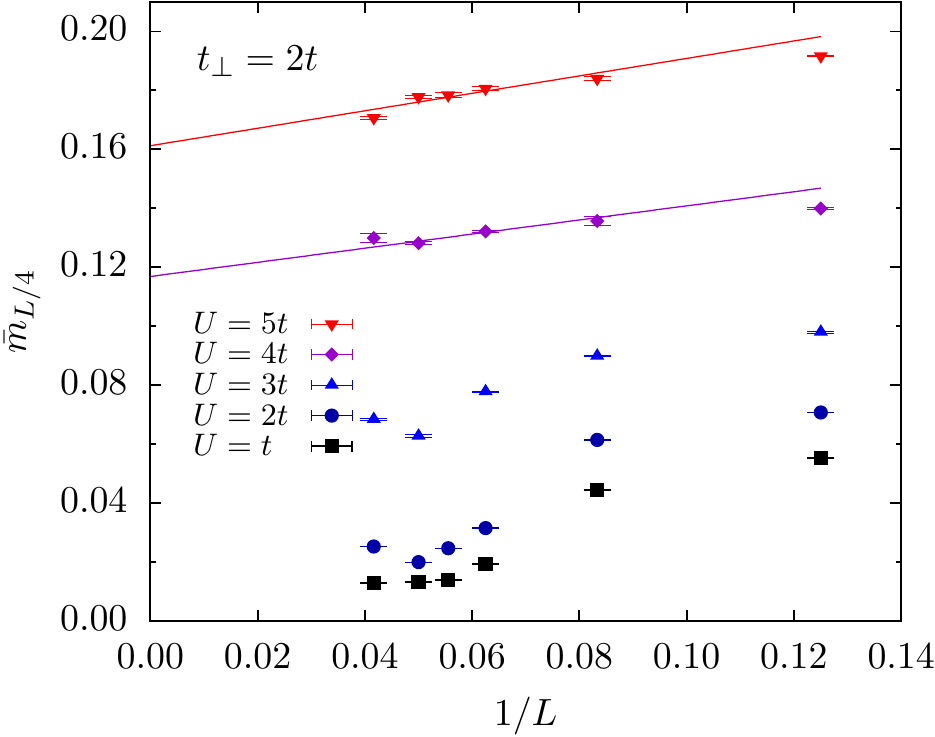}
  \caption{(color online) Finite size scaling of the staggered magnetization for $\tp=t$ (top) and $\tp=2t$ (bottom). For couplings $U/t\geq 4$, linear fits were performed (solid lines).}
\label{fig:tp2mag}
\end{figure}

The results for $\bar m_{L/4}$ are shown in Fig.~\ref{fig:tp2mag} as a function of the inverse system size $1/L$ for interlayer hoppings $\tp=t$ and $2t$. 
At this point a finite-size scaling analysis needs to be performed, e.g.~in terms of  a polynomial extrapolation to the thermodynamic limit at $1/L=0$ to access a thermodynamic limit estimate  for the staggered magnetization. This procedure works reasonably well at couplings $U \gtrsim 4t$, for which a linear extrapolation yields robust finite estimates for the staggered magnetization, thus indicating an emerging AF state. However, for weaker interactions, the finite-system estimates for the staggered magnetization exhibits a strong nonmonotonous behavior upon varying the system size, that defies a similarly reliable extrapolation.

\begin{figure}[t!]
  \centering
\includegraphics[width=\columnwidth]{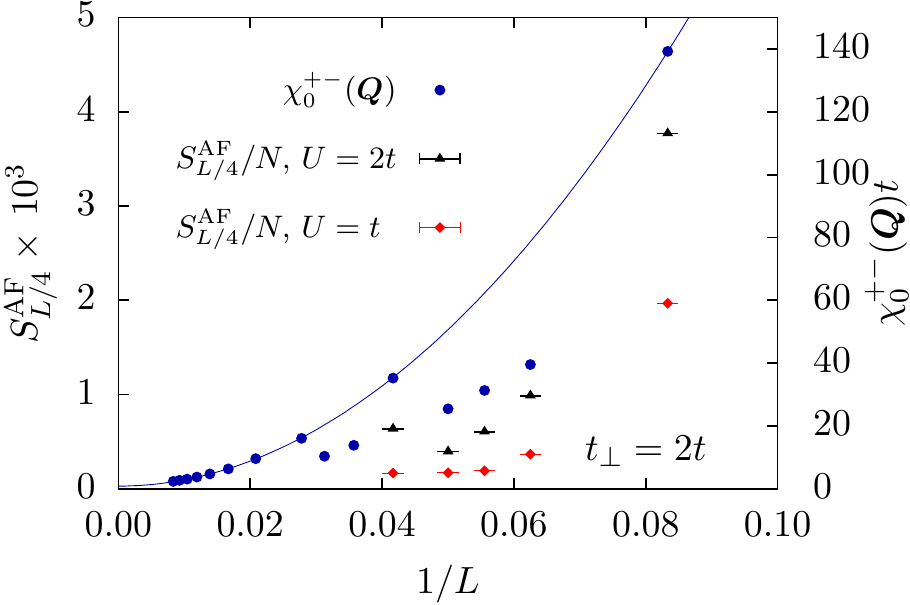}
  \caption{(color online) Comparison of the antiferromagnetic structure factor of the weakly interacting system and the noninteracting
    static susceptibility $\chi^{+-}_0(\bm Q)$ at finite temperature $T=10^{-4}t$ (blue dots) for different linear lengths $L$. A quadratic function may be fitted to the susceptibility data for lengths $L$ that are multiples of 12 (blue solid line).}  
\label{fig:tp2chi}
\end{figure}

This peculiar finite-size behavior can be traced back to the coarse sampling of the Fermi surface for small finite systems. In fact, a similar nonmonotonous dependence on the system size can be observed already in the noninteracting system's static spin-spin susceptibility, Eq.~(\ref{eq:suscep}):
While at $T=0$ the susceptibility diverges at the nesting vector $\bm Q$, at low finite temperatures it exhibits a very similar finite-size dependence as the one observed for the AF structure factor at small finite values of $U$. This is evident from Fig.~\ref{fig:tp2chi}, where both quantities are compared for the case of $\tp=2t$. From the behavior of $\chi^{+-}_0(\bm Q)$ it appears that if only system sizes with $L$ being multiples of 12 are considered, a quadratic behavior can be well fitted to the finite-size susceptibility data, yielding a finite estimate in the thermodynamic limit. In order to perform a similar meaningful finite size scaling for the finite-$U$ structure factor data, results for system lengths $L=36,48,\dots$ would be required, which are currently out of reach. On the other hand, the low-$U$ data for $\bar m_{L/4}$ in Fig.~\ref{fig:tp2mag} 
may also be argued to exhibit a leveling-off behavior towards a small, finite value at the lower range of $1/L$ values. 
While thus no definite statement can be made based on our data about the size of the staggered magnetization at the lower values of $U$ in the thermodynamic limit, 
the data in Fig.~\ref{fig:tp2mag} is consistent with the persistence of a small, but finite value of the staggered magnetization in the thermodynamic limit for both $U=t$ and $U=2t$, which would be in accord with the weak coupling scenario  from the fRG analysis.  
Previous DQMC simulations\cite{bouadim2008} suggested the existence of a paramagnetic metallic phase at small values of $U/t$; however, the corresponding calculations were performed at low but finite temperatures and were restricted to linear system sizes $L\leq10$ only. Based on our ground state analysis with $L$ up to 24, we cannot confirm the persistence of the
 paramagnetic phase. Our DQMC data are also consistent with the (more conventional) scenario supported by the fRG calculations, restricting the range of the paramagnetic metallic phase down to the $U=0$ line.

\begin{figure}[t!]
  \centering
\includegraphics[width=\columnwidth]{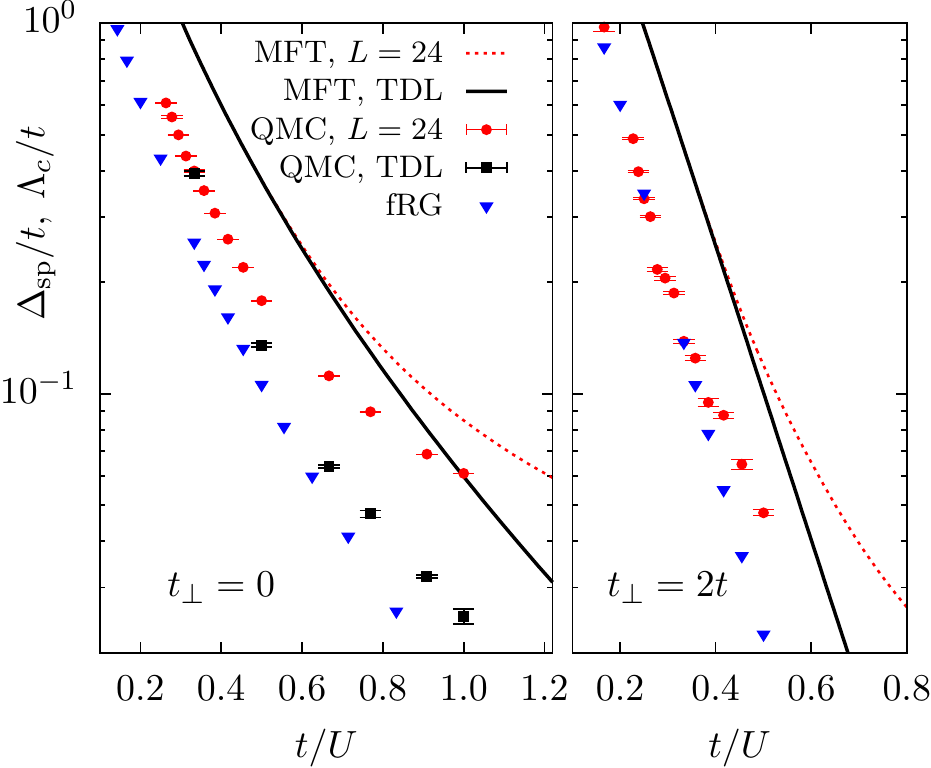}
  \caption{(color online) Comparison of the single-particle gaps from HFMFT (solid and dashed lines), QMC (red dots and black squares) and the critical scale from fRG (blue triangles). Left: Single-layer case. Right: Bilayer system with $t_\perp=2t$.}  
\label{fig:gaps}
\end{figure}

Within this scenario, the AF Mott insulating phase is accompanied by a finite single-particle gap. Based on the DQMC calculations, we can extract this excitation gap from the large-imaginary time 
limit of the Matsubara Green's function,
\begin{equation}
  \label{eq:green-tau}
  G_{\bm k}(\tau)=-\langle c^{}_{\bm k}(\tau)c^\dagger_{\bm k}(0)\rangle\stackrel{\tau\to\infty}{\propto} e^{-\Delta(\bm k)\tau}.
\end{equation}
The single particle gap $\Delta_{\rm sp}$ is then defined as the smallest occurring value, $\Delta_{\rm sp}=\min\{ \Delta(\bm k)\}$.
This quantity allows for a direct comparison to the gap obtained within HFMFT as well as the critical scale obtained from the fRG calculations. 
In Fig.~\ref{fig:gaps}, the corresponding results from the three different methods are compiled for the single layer system (left panel) and the bilayer system at $t_\perp=2t$ (right panel). Generally, QMC and fRG results indicate the exponential behavior obtained within the HFMFT calculations. The deviations that are observed at weak coupling in the finite-size QMC data are also obtained within  HFMFT calculations on such finite systems. This indicates again the presence of sizable finite-size effect in the low-$U$ regime.
For the interacting single-layer system, the immediate onset of antiferromagnetism at any finite $U$ is by now well established~\cite{halboth2000,varney2009,schaefer2014}.
Even though a reliable extrapolation of the QMC data is limited to couplings $U\geq 2t$ for the Hubbard bilayer, the comparison in Fig.~\ref{fig:gaps} supports the validity of the fRG scenario in the weak coupling regime and thus makes a further case for the emergence of the AF Mott insulator phase for any weak, finite interaction strength from the free metallic region.

%%%%%%%%%%%%%%%%%%%%%%%%%%%%%%
\section{Antiferromagnet to Band Insulator Transition}
\label{sec:afbi}
%%%%%%%%%%%%%%%%%%%%%%%%%%%%%%
As already discussed in Sec.~\ref{sec:tb-hf}, the free system $(U=0)$ undergoes a transition from the metal to a band insulator at $t_\perp=4t$. 
On the other end of the interaction range, i.e. in the strong coupling limit $U\to\infty$, the low-energy spin dynamics in the bilayer Hubbard model can be effectively described by a spin-$1/2$ Heisenberg model,
\begin{equation}
  \label{eq:heisenberg}
  H=J\sum_{\langle ij\rangle\lambda}\bm S_{i\lambda}\cdot\bm S_{j\lambda}+J_\perp\sum_{i}\bm S_{i1}\cdot\bm S_{i2},
\end{equation}
with antiferromagnetic couplings $J=4t^2/U$ and $J_\perp=4t_\perp^2/U$. 
For this spin model, a quantum phase transition from an antiferromagnet low-$J_\perp$ phase to a magnetically disordered dimerized phase with spin singlets predominantly formed on the interlayer bonds occurs at $J_\perp/J=2.52181(3)$~\cite{hida1992,millis1993,sandvik1994,wang2006}.
With respect to the Hubbard model parameters, this corresponds to a critical interlayer hopping of $t_\perp/t=1.58802(1)$.
As shown in Fig.~\ref{fig:phase-diag}, this quantum disordered, dimerized phase connects adiabatically to the $U=0$ band insulator phase, while the AF Mott insulator is separated from the large-$t_\perp$ band insulator regime by a magnetic quantum phase transition related to the breaking of the spin rotation symmetry in the AF phase.  
In this section, we discuss how the extent of the AF phase can be extracted from DQMC and fRG for finite values of the interaction strength $U/t$. 
\subsection{Quantum Monte Carlo}
\label{sec:qmc-afbi}
\begin{figure}[t!]
  \centering
  \includegraphics[width=\columnwidth]{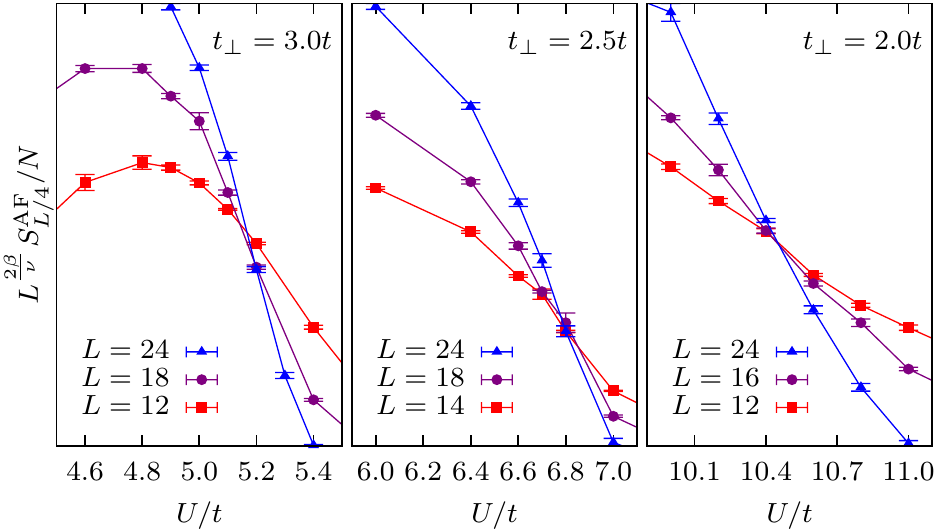}
  \caption{(color online) Finite size scaling analysis of the rescaled antiferromagnetic structure factor around the AF-BI transition, assuming $\beta=0.369$ and $\nu=0.711$. The crossing points indicate the critical coupling $U_c/t$ of the transition.}
  \label{fig:criticalUs}
\end{figure}

\begin{figure}[t]
  \centering
  \includegraphics[width=\columnwidth]{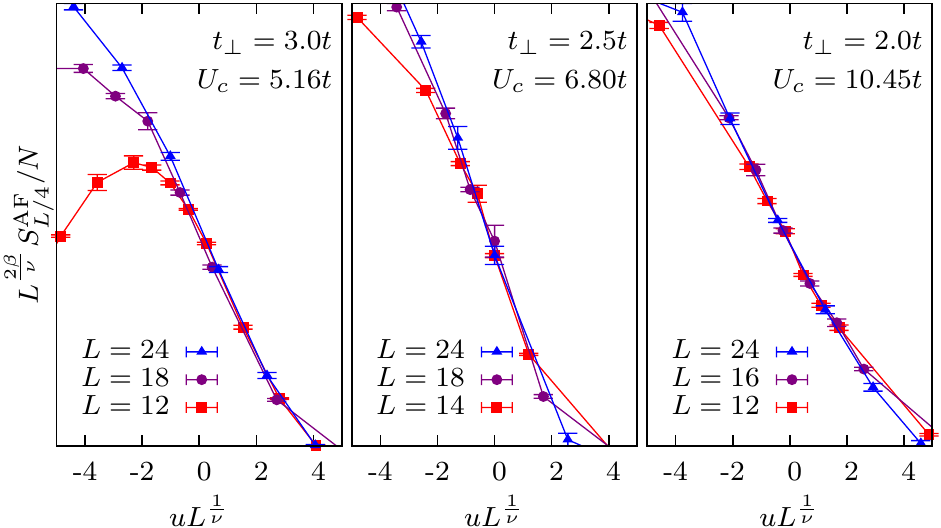}
  \caption{(color online) Data collapse analysis of the rescaled antiferromagnetic structure factor around the AF-BI transition, assuming $\beta=0.369$ and $\nu=0.711$.}
  \label{fig:collapse}
\end{figure}

We  employ finite size scaling of the DQMC structure factor data to locate the transition between the Mott-insulating phase and the band-insulating regime. In the following, we locate the transition points upon varying $U$ at fixed values of $t_\perp/t$. 
Given the limits to be $t_\perp=4t$ for $U=0$ and $t_\perp=1.588t$ for $U\to\infty$, we extract the critical coupling $U_c$ within the intermediate interlayer hopping range at $t_\perp/t=2.0, 2.5$, and $3.0$.
In order to determine $U_c$, we make use of finite size scaling with the following standard scaling ansatz for the structure factor in the proximity of the quantum phase transition:
\begin{equation}
  \label{eq:scaling}
  S^{\rm AF}/N=L^{-\frac{2\beta}\nu}{\cal F}_S\big(uL^{\frac1\nu}\big), \quad u=\frac{U-U_c}{U_c},
\end{equation}
where ${\cal F}_S$ is the scaling function of the structure factor, and $\beta$, $\nu$ denote the critical exponents for the order parameter and the correlation length, respectively. At the critical coupling $u=0$, the scaling function will be evaluated at the same point for different $L$, so that the rescaled structure factors $L^{\frac{2\beta}\nu} S^{\rm AF}/N$ should cross at $U=U_c$. Thus, by monitoring this observable over a range of interactions $U/t$, curves for different $L$ should intersect at the critical coupling $U_c/t$. Since the onset of  AF order in the ground state of this two-dimensional quantum system breaks the $SU(2)$ spin rotational symmetry, we anticipate that the quantum phase transition belongs to the universality class of the three-dimensional classical Heisenberg model\cite{chakravarty1988,haldane1988}. We thus employ the critical exponents of this universality class $\beta=0.3689(3)$ and $\nu=0.7112(5)$\cite{campostrini2002}. From the corresponding analysis shown in Fig.~\ref{fig:criticalUs}, we extract for the critical line $U_c(\tp/t)$ the three points $U_c(3.0)=5.16(5)t$, $U_c(2.5)=6.80(5)t$ and $U_c(2.0)=10.45(5)t$. The estimated phase boundary points have already been indicated in Fig.~\ref{fig:phase-diag}. The data collapses shown in Fig.~\ref{fig:collapse} demonstrate, that the finite-size data indeed follows the anticipated scaling ansatz in Eq.~(\ref{eq:scaling}) within the vicinity of the quantum critical points. Fig.~\ref{fig:collapse} also exhibits an increase of further finite-size corrections for lower values of the critical interaction strength, in accord with our findings in Sec.~\ref{qmc}.

\begin{figure}[t]
  \centering
  \includegraphics[width=\columnwidth]{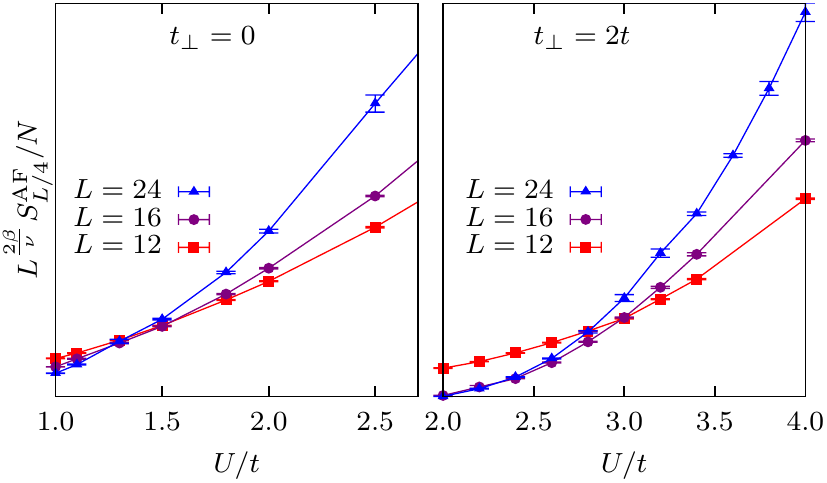}
  \caption{(color online) Apparent crossings of the rescaled structure factor data for $t_\perp=0$ (left panel) and $t_\perp=2t$ (right panel) in the lower-$U$ region, indicative of the crossover in the nature of the AF state.}
  \label{fig:crossover}
\end{figure}

At this point, it is  interesting to observe, that the enhanced susceptibility  towards the free system's Fermi surface structure at low values of $U$ in effect leads to a change in the finite-size scaling behavior of the structure factor $S^{\rm AF}$. This is seen from e.g.~the right panel of Fig.~\ref{fig:crossover}  at $t_\perp=2t$, which could 
be mistaken to indicate a magnetic phase transition  near $U/t\approx 3$. These apparent crossings instead signal a crossover in the characteristics of the AF state: while at low values of $U$, 
the AF order arises from a Fermi surface instability, it relates for larger values of $U$ to the localized nature of the Mott insulator with the emerging Heisenberg spin physics. In fact, a 
similarly misleading indication for such a transition is observed in the decoupled layer ($t_\perp=0$) data, shown in the left panel of the same figure, near $U/t\approx 1.5$.
From our DQMC simulations, we thus find that on the bilayer, this crossover behavior is even more pronounced than for the single-layer model, due to the more complex Fermi surface structure in the bilayer case and the corresponding suppression of the van Hove singularity in the DOS away from the  half-filled system.

%
%%%%%%%%%%%%%%%%%%%%%%%%%%%%%%%%%%
\subsection{Functional Renormalization Group}
\label{sec:frg-afbi}
%%%%%%%%%%%%%%%%%%%%%%%%%%%%%%%%%%
It is interesting to assess, if and how the transition out of the AF state upon increasing $U$ can be identified within the fRG approach. 
\begin{figure}[t!]
\centering
   \includegraphics[height=0.9\columnwidth]{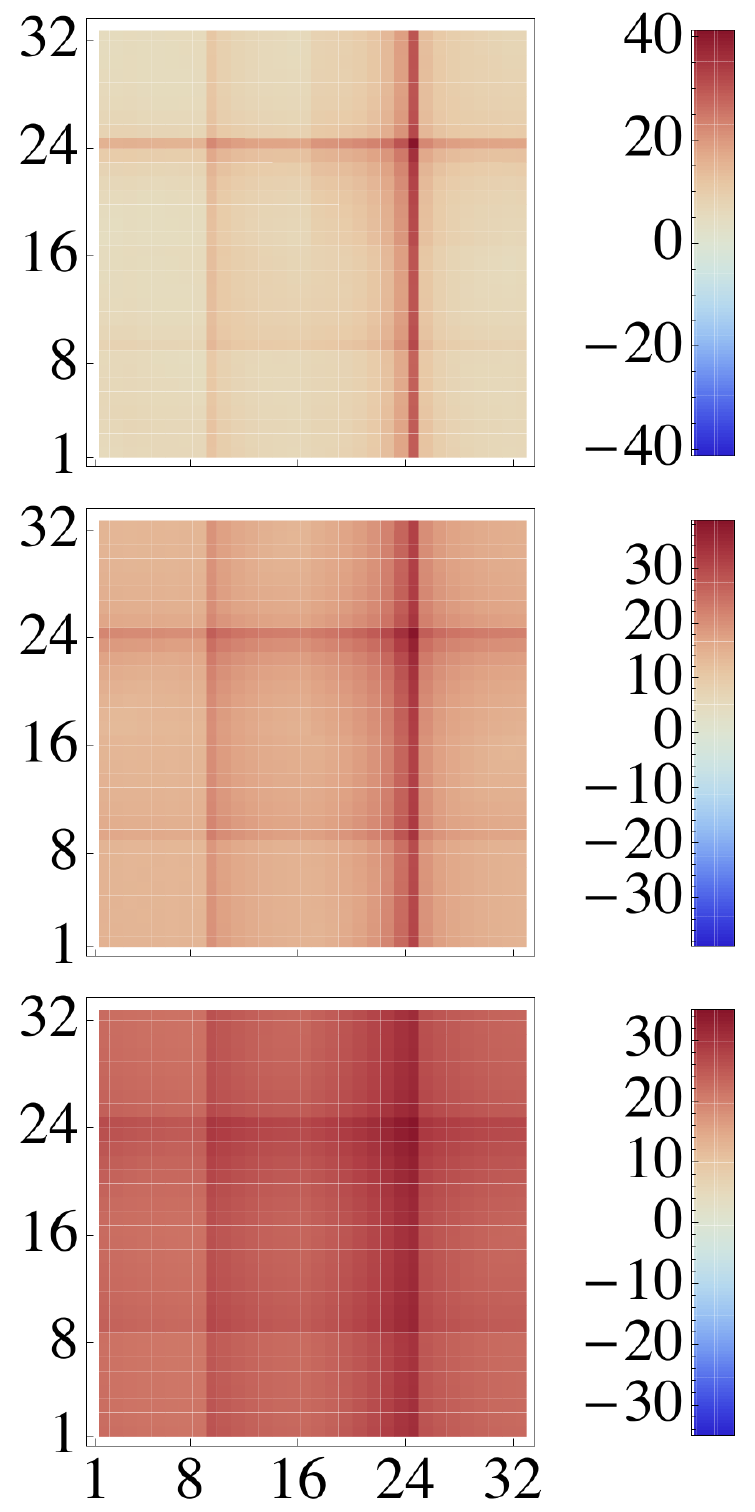}
   \hspace{0.3cm}
            \includegraphics[height=0.92\columnwidth]{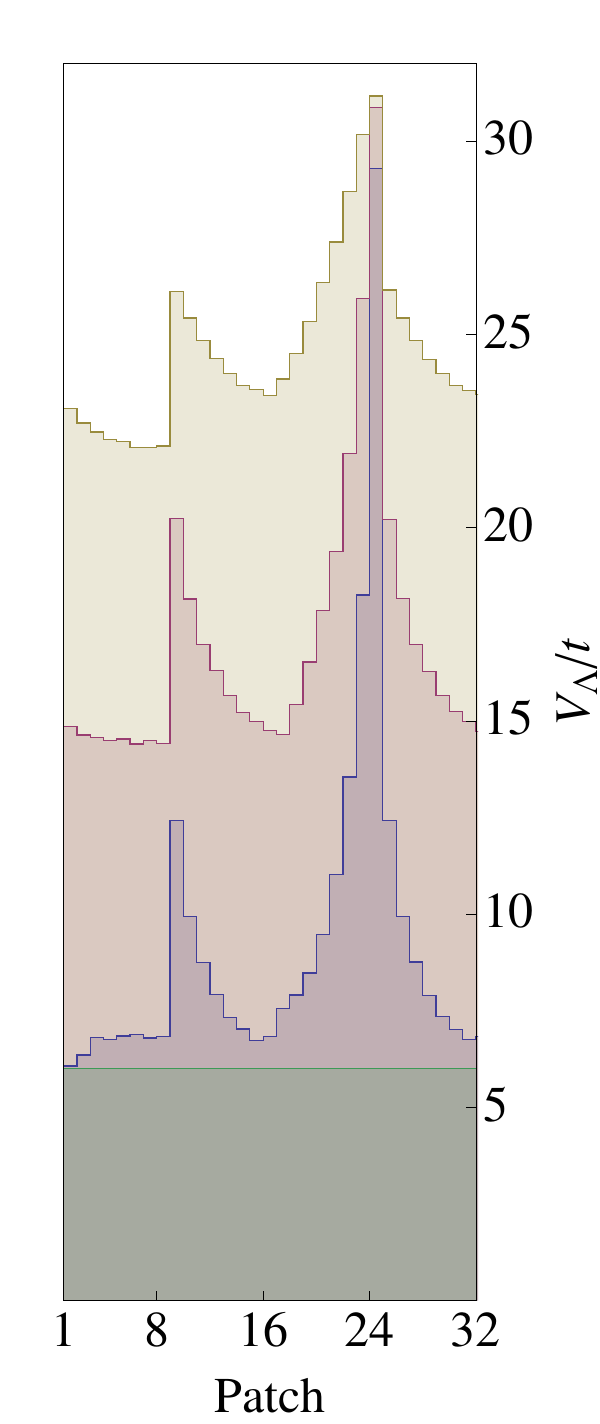}
  \caption{(color online) left panel: Effective low-energy interaction vertex in units of $t$ for a combination of layer indices $\lambda_i$ where $\lambda_1=\lambda_2=\lambda_3=\lambda_4$ for three choices of $t_\perp \in\{2t,3.5t,3.9t\}$ (from top to bottom) at fixed onsite interaction $U=6t$. We present the vertex at an RG scale $\Lambda^*$ where its largest component has grown large, $V_{\Lambda,\text{max}}\sim 30t$. For these combinations of parameters we find $\Lambda^* \sim\mathcal{O}(0.5t)$. The conventions are identical to the ones in Fig.~\ref{fig:fRGvertex}. Right panel: Interaction vertex profile for fixed $k_1$ at patch 1 for the $t_\perp$ from the left panel. This corresponds to a horizontal cut of the interaction vertices shown in the left panel. The flat line in the bottom (green) shows the value of the initial interaction.}
  \label{frgBIMI}
\end{figure}
The fRG method is a controlled approximation for weak interactions and has been shown to provide reliable results also in the intermediate interaction regime. 
Here, we thus study the behavior of the bilayer system over a range of $U/t$ values from small interactions up to the order of the bandwidth. 
For small interlayer hopping $t_\perp/t$ and at any investigated value of the onsite interaction $U$, we observe the appearance of the characteristic feature of a clear AF instability in the interaction vertex, which is sharply pronounced in momentum space, as shown in Fig.~\ref{fig:fRGvertex}. 
\begin{figure}[t!]
\centering
   \includegraphics[width=1.0\columnwidth]{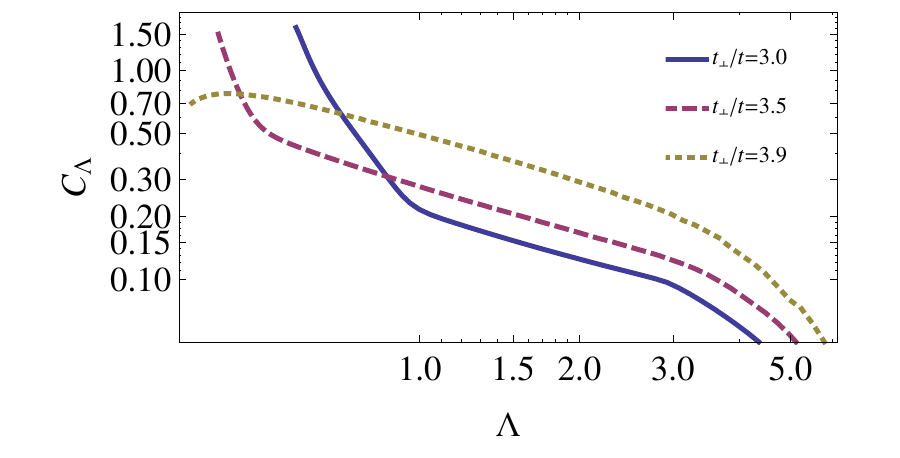}
  \caption{(color online) Flow of the inverse width $C_\Lambda$ for three choices of $t_\perp \in\{3t,3.5t,3.9t\}$ (solid, dashed and dotted line, respectively) at fixed onsite interaction $U=6t$.}
  \label{cflow}
\end{figure}
As explained in Sec.~\ref{sec:frg_weak}, this specific vertex structure leads to the identification of the AF insulator state. 
Interestingly, 
for larger values of $t_\perp/t$, we observe, as shown in Fig.~\ref{frgBIMI}, that these sharp structures in the vertex $V_\Lambda$ smear out as we increase $U\gtrsim 4t$. 
This smearing happens gradually, and for fixed onsite interaction beyond a value of $t_\perp$ smaller than $4t$ the characteristic AF structure vanishes completely from the vertex function, cf. Fig.~\ref{frgBIMI}.
The sizeable values of the onsite interactions investigated here, are usually expected to lie beyond the regime of applicability of the presented fRG approach, operating on this level of truncation.
However, in light of the above results from the DQMC simulations, we may interpret the above behavior also as a signature for the  breakdown of the AF order and use this to obtain an fRG estimate for the transition region towards the band insulating state.

Therefore, we resort to an effective parameterization of the two-particle interaction vertex $V_{\Lambda}$ by using an exchange propagator $\propto 1/(C_{\Lambda} {\bf q}^2+m_{\Lambda})$ following the idea of a gradient expansion around ordering momenta\cite{samaier2014}. 
The inverse of the maximum interaction component determines $m_{\Lambda}=1/V_{\Lambda ,\mathrm{max}}$ and the inverse width $C_{\Lambda}$ is related to the spin stiffness. 
From the flow of $C_{\Lambda}$, we then read off the AFMI to BI transition, see Fig.~\ref{cflow}: 
In the AF regime $C_{\Lambda}$ grows large corresponding to a sharp momentum structure or a long-ranged correlation in position space.
In contrast, the flow of $C_{\Lambda}$ stays finite in the band insulating phase, where the correlations become short-ranged leading to dimers accross the two layers. 
The corresponding transition line is presented in Fig.~\ref{fig:phase-diag}.
The above parameterization is reasonable in the regime where antiferromagnetic correlations become important and otherwise is less reliable.
In this way, transition line should only be taken as a rough estimate.
Nevertheless, from our fRG data, we may thus ascertain the disappearance of the AF instability for sizable values of $U$ and $t_\perp$ as expected for the Mott-insulator to band-insulator transition and we may determine the approximate position of the transition region from the above procedure. 
Given the expected less quantitative accuracy of the fRG approach in this coupling range, we consider the transition as extracted above from the fRG to agree reasonably well to the DQMC transition points, cf.~Fig.~\ref{fig:phase-diag}.

The above analysis reveals that the fRG approach allows to detect a softening of the AF order such as near the transition to the band insulator phase. Hence, we take the absence of any similar weakening in the AF structure of the interaction vertex in the low-$U$ region as another argument against the relevance of interlayer singlet formation as a mechanism to stabilize a low-$U$ paramagnetic metallic state.

%%%%%%%%%%%%%%%%%%%%%%%%%%%%%%%%%%
\section{Conclusions}
%%%%%%%%%%%%%%%%%%%%%%%%%%%%%%%%%%
\label{concl}
Based on a combined analysis using fRG and DQMC calculations, we established the nature of the quantum phase diagram of the half-filled Hubbard model on the square lattice bilayer. In particular, we identified the dominant AF instability in the weak-coupling region as a Fermi surface instability of the metallic free state. Within both the fRG and the DQMC simulations, we did not detect any competing instabilities at weak-coupling. Based on a careful analysis of the ground state DQMC data for the staggered structure factor, we identified the main difficulty in performing a standard finite-size extrapolation in the weak-coupling region, due to an enhanced susceptibility of the magnetic state on the finite system size, which relates to the complex Fermi surface structure in the free metallic region. Our DQMC data is in accord with the HFMFT and our fRG scenario in which the metallic region is restricted to the $U=0$ line, reflecting the persistent perfect nesting in the bilayer lattice. Furthermore, we  employed standard finite size scaling to trace out within DQMC the transition line between the Mott insulator and the band-insulating regime, and also obtained  signatures for this transition from a careful analysis of the fRG interaction vertex flow. For the future, it will be interesting to address the stability of the AF phase with respect to extended interactions both within the planes as well as upon the addition of interlayer interactions.

%%%%%%%%%%%%%%%%%%%%%%%%%%%%%%%%%%
%{\bf{\it{Acknowledgements}}}\\
%%%%%%%%%%%%%%%%%%%%%%%%%%%%%%%%%%
%
\acknowledgments

We acknowledge discussions with Fakher F.~Assaad, Stephan Haas, Carsten Honerkamp, Stefan A.~Maier, Jan M.~Pawlowski, and  Daniel D.~Scherer. L.C. acknowledges support by the HGSFP. M.M.S. is supported by the grant ERC-AdG-290623 and DFG grant FOR 723. M.G. and S.W. acknowledge support from the DFG within grant FOR 1807 and grant WE 3949/3-1. Furthermore, we acknowledge the allocation of CPU time through JARA-HPC at JSC J\"ulich and at RWTH Aachen. 

%%%%%%%%%%%%%%%%%%%%%%%%%%%%%%%%%%
\appendix
%%%%%%%%%%%%%%%%%%%%%%%%%%%%%%%%%%

%%% Hartree-Fock-results
\section{Bilayer in Hartree-Fock approximation}
\label{sec:hubb-mf}
We show in this appendix that the half-filled bilayer system in HFMFT approximation
is antiferromagnetically ordered for any finite value of $U/t$ for $t_\perp<4t$. In HFMFT,  
the interaction term is decoupled as
\begin{equation}
  \label{eq:hf-decoupling_app}
  n_{i\lambda\up}n_{i\lambda\down}\rightarrow n_{i\lambda\up}\avg{n_{i\lambda\down}}+\avg{n_{i\lambda\up}}n_{i\lambda\down} -\avg{ n_{i\lambda\up}}\avg{ n_{i\lambda\down}}
\end{equation}
We now assume a staggered magnetization $m$ on the sublattices
$\sigma=A,B$. We divide the lattice in $N_C=\frac{N}{4}$ unit cells with four sites, labeled by $(\lambda,\sigma)$,
and make the following ansatz for the mean field parameters of the AF state:
\begin{align}
  \label{eq:mfansatz}
  \avg{n_{1A\up}}=\tfrac12 +m&&\avg{n_{1A\down}}=\tfrac12-m\\
  \avg{n_{1B\up}}=\tfrac12 -m&&\avg{n_{1B\down}}=\tfrac12+m\\\nonumber
  \avg{n_{2A\up}}=\tfrac12 +m&&\avg{n_{2A\down}}=\tfrac12-m\\\nonumber
  \avg{n_{2B\up}}=\tfrac12 -m&&\avg{n_{2B\down}}=\tfrac12+m\nonumber
\end{align}
This, after Fourier transformation via $c^{}_{i\lambda \sigma s}=\sqrt{1/N_C}\sum_{\bm k}e^{-i\bm k\bm r_i} c^{}_{\bm k \lambda \sigma s}$ yields the following HFMFT Hamiltonian
\begin{align}
  \label{eq:hf-ham}
  H^{\rm HF}=&\sum_{\bm k s}\bm c^\dg_{\bm k s} M(\bm k)\, \bm c_{\bm k s} + NUm^2+\frac{NU}2,\\
  \bm c_{\bm k s}=&\begin{pmatrix}c_{\bm k 1A s}\\c_{\bm k 1B s}\\c_{\bm k 2A s}\\c_{\bm k 2B s}\end{pmatrix},\,M(\bm k)= \begin{pmatrix}
    Um&\epsilon_0(\bm k)&0&-\tp\\ \epsilon_0(\bm
    k)&-Um&-\tp&0\\0&-\tp&Um&\epsilon_0(\bm k)\\-\tp&0&\epsilon_0(\bm k)&-Um \end{pmatrix}.\nonumber
\end{align}
Here, $\epsilon_0(\bm k)=\epsilon^+_0(\bm k,0)$ denotes the square lattice single-layer dispersion.
The matrix $M(\bm k)$ is independent of the spin index $s$, and upon diagonalization,
we arrive at four bands $\alpha\in\{1,...,4\}$,
\begin{equation}
  \label{eq:hf-disp-app}
  \epsilon^\alpha_{\rm HF}(\bm k,\tp)=\pm\sqrt{(Um)^2+\left(\epsilon^\pm_0(\bm k,\tp)\right)^2}.
\end{equation}
Since $Um$ corresponds to (half) the enery gap between the upper and lower bands, we define the gap parameter $\Delta=Um$.
The HF energy density $E(\Delta)=\avg{H^{\rm HF}}/N$ is then expressed in terms of the gap as
\begin{align}
\label{hf-energy}
  E(\Delta)&=\frac{1}{N}\sum_{\bm k \alpha s}f\big(\epsilon^\alpha_{\rm HF}(\bm k,\tp)\big)\,\epsilon^\alpha_{\rm HF}(\bm k,\tp) +\frac{\Delta^2}{U}\\
&=-2 \int^{\infty}_{0}\! d\epsilon\, \rho(\epsilon,\tp)\sqrt{\Delta^2+\epsilon^2} +\frac{\Delta^2}{U},\nonumber
\end{align}
where $f(\epsilon)$ denotes the Fermi-Dirac distribution. To obtain the second line of Eq.~(\ref{hf-energy}), we set $T=0$ and exploited the particle-hole symmetry of the system.
In order to find the critical interaction $U_c/t$, we expand the
HF energy in $\Delta$:
\begin{equation}
  \label{eq:ener-exp}
  E(\Delta)=E(0)+\frac{\Delta^2}{U}(1-U\chi_0)+{\cal O}(\Delta^4), \;\chi_0=\int^\infty_0\!\!\! d\epsilon\,\frac{\rho(\epsilon,\tp)}{\epsilon}
\end{equation}
The AF state will be favorable in energy, if the second order term becomes negative,
i.e., at 
$U_c=\chi_0^{-1}$. Thus, in the case of a diverging $\chi_0$, any finite $U$ leads to an AF
instability, i.e., $U_c=0$. And indeed, for a finite DOS at $\epsilon=0$, $\chi_0$ exhibits a logarithmic divergence, due
to the $1/\epsilon$ singularity of the integrand at $\epsilon=0$.

Finally, we take a closer look at the functional behavior of the gap $\Delta(U)$.
By demanding $\frac{\partial E}{\partial\Delta}=0$, we arrive at the gap equation:
\begin{equation}
  \label{eq:hf-gapeq}
  1=U\int^\infty_0\!d\epsilon\,\frac{\rho(\epsilon,\tp)}{\sqrt{\Delta^2+\epsilon^2}}.
\end{equation}
In the single-layer system ($t_\perp=0$), the gap equation can be approximately solved~\cite{hirsch1985} and exposes a modified exponential behavior $\Delta\sim t\exp(-2\pi\sqrt{t/U})$, while for finite values of $t_\perp$, a numerical solution reveals a conventional exponential scaling $\Delta\sim t\exp(-C_{\tp} t/U)$, cf.~Fig.~\ref{fig:gaps}. The positive coefficient $C_{\tp}$ increases continuously from $C_{0.1t}=3.88$ to $C_{3.9t}=12.4$, which upon approaching $U\to 0$ results in a faster decay of the gap with increasing interlayer hopping.

% \begin{figure}[t!]
%   \centering
%   \includegraphics[width=\columnwidth]{mfgaps}
%   \caption{Solution of the Hartree-Fock gap equation for different interlayer hoppings $\tp/t$. }
%   \label{fig:hfgap}
% \end{figure}

%%%%%%%%%%%%%%%%%%%%%%%%%%%%%%%%%%

\end{document}